# Fully Convolutional Spatio-Temporal Models for Representation Learning in Plasma Science

Ge Dong[1(a)], Kyle Gerard Felker[2], Alexey Svyatkovskiy[3], William Tang[1*], Julian Kates-Harbeck[1*]

[1] Princeton Plasma Physics Laboratory, Princeton, NJ
[2] Argonne National Laboratory, Lemont, IL
[3] Microsoft Corporation, Redmond, WA
*joint supervision

[(a)]gdong@princeton.edu

Abstract:

We have trained a fully convolutional spatio-temporal model for fast and accurate representation learning in the challenging exemplar application area of fusion energy plasma science. The onset of major disruptions is a critically important fusion energy science (FES) issue that must be resolved for advanced tokamak plasmas such as the $25B burning plasma ITER experiment. While a variety of statistical methods have been used to address the problem of tokamak disruption prediction and control, recent approaches based on deep learning have proven particularly compelling. In the present paper, we introduce further improvements to the fusion recurrent neural network (FRNN) software suite, which delivered cross-machine disruption predictions with unprecedented accuracy using a large database of experimental signals from two major tokamaks. Up to now, FRNN was based on the long short-term memory (LSTM) variant of recurrent neural networks to leverage the temporal information in the data. Here, we implement and apply the "temporal convolutional neural network (TCN)" architecture to the time-dependent input signals, thus rendering the FRNN architecture fully convolutional. This allows highly optimized convolution operations to carry the majority of the computational load of training, thus enabling a reduction in training time, and the effective use of high performance computing (HPC) resources for hyperparameter tuning. At the same time, the TCN based architecture achieves equal or better predictive performance when compared with the LSTM architecture for a large, representative fusion database. Across data-rich scientific disciplines, these results have implications for the resource-effective training of general spatio-temporal feature extractors based on deep learning. Moreover, this challenging exemplar case study illustrates the advantages of a predictive platform with flexible architecture selection options capable of being readily tuned and adapted for responding to prediction needs that increasingly arise in large modern observational dataset.

I. Introduction

Deep learning has become an increasingly important methodology for the effective analysis and interpretation of big data in modern social and scientific areas [1-3]. In this study, we discuss the application of deep learning models in the prominent exemplar problem of disruption predictions in tokamaks [4], which are magnetic fusion experimental devices with large numbers of advanced diagnostics to monitor spatio-temporal plasma performance.

In many toroidal plasma devices such as tokamaks and spherical toruses (ST's) [5], disruptions are observed as sudden and dangerous events that induce rapid release of particles and energy to the device wall [6]. A typical disruption brings the plasma experiment (the "shot") to an abrupt end and, because of the associated large rapid energy release, it can also seriously damage the device – especially in larger systems such as ITER [7]. Accordingly, the development of a plasma control system (PCS) with the ability to reliably detect and subsequently mitigate or avoid the majority of the disruption events [8] is broadly regarded as the most important milestone for establishing the viability of future larger tokamak devices to deliver a fusion energy reactor.

To robustly mitigate or prevent disruptions, the first step for the PCS is to accurately predict disruptions as early as possible. Traditional methods for disruption studies and predictions range from using simple empirical formulae and analytic expressions to more sophisticated first-principles-based simulations, such as the magnetohydrodynamic (MHD) models [9] and the gyrokinetic models [10]. These simulations are used to study the dynamics and mechanisms of disruptions, and accordingly to advance their prediction and also their possible avoidance through active plasma control. Empirical expressions and even MHD models are generally simple and fast enough to be implemented in the PCS and can thereby aid real-time disruption predictions and control. However, these models often contain insufficient physics information to correctly predict complex or novel scenarios, including device operations for different parametric and/or hardware regimes (such as those involving, for example, reactor-relevant wall materials [11]). In order to better capture nonlinear dynamics and physics associated with realistic magnetic geometry, large-scale first-principles-based simulations can be engaged. However, the vast computational resources required to allow real-time predictive capabilities for such simulations make an implementation in the PCS infeasible. Consequently, complementary to the two aforementioned categories of models, emerging big-data-driven methodologies have become an increasingly powerful modern approach addressing the grand challenge of prediction and control of tokamak disruptions [12]. We have accordingly developed a deep learning capability of general computational science interest that is capable of enabling significant progress toward resolving this major application science exemplar problem via effective utilization of the hardware capabilities of modern leadership class supercomputing systems. This involves the training of multiple models with distinct architectures within a single software suite adaptable to different temporal and spatial learning tasks – with natural connections to enabling ensemble schemes for highly accurate prediction.

We focus on and refer to models that can be easily built and extended in a layered fashion – as well as optimized (trained) using automatic differentiation and back-propagation (such as the neural networks) – as "deep learning" models. Machine learning models that do not have this property are referred to as "classical" models. Examples include support vector machines [13] and random forest algorithms [14], both of which have been applied successfully to advance disruption prediction capability. So far, a key advantage of deep learning models has been their ability to perform cross-machine predictions – forecasting the plasma behavior in an experiment never seen during training and or validation. This is especially key for establishing the relevance of such studies for ITER, as ITER will not be able to withstand enough disruptions for a large training data set [15].

Our starting point in the present paper begins with a careful examination of the machine learning framework of the fusion recurrent neural network (FRNN) first introduced in [4]. We proceed to move forward with the design and implementation of a temporal convolutional neural network (TCN) [16] most suitable for the temporal representation of the extensive database of FES input signals of interest [4]. The associated TCN architecture based on dilated causal convolutions [17] has some advantages compared with the long short-term memory (LSTM) architecture in FRNN [4]. The TCN architecture will be introduced in detail in the next section.

Comparisons between TCN and LSTM based architectures for various fusion databases are described in detail in Section III of this paper. While TCN based architectures generally achieve superior predictive power and improved computational performance, we note that for certain tasks and databases, the original LSTM-based model [4] can still outperform all alternatives. From a general computational science perspective, this finding highlights the importance of maintaining multiple architectures in a modern software suite such as FRNN. Moreover, the choice of temporal processing layer (such as LSTM cell versus TCN) can be viewed as a high level architectural hyperparameter for FRNN. This framing aligns with our vision for an AI/DL based platform with flexible and adaptive model architectures that can be automatically hyperparameter tuned for various tasks and databases associated, for example, with future fusion plasma predictions and analysis tasks. This in turn has implications for building capabilities to face future real-time plasma control challenges.

II. Model Architecture

Figure 1 shows the schematic of the new deep learning architecture based on TCN's introduced here into FRNN. As in the previous LSTM based models [4], the TCN-based FRNN models process inputs composed of two main types of signals: 0D scalar signals (such as the plasma current), and 1D profile signals (such as the electron temperature profile). The descriptions and numerical properties of the fourteen 0D and two 1D signals can be found in Table 1. Example time series of signals available on DIII-D are shown in the top four panels of Figure 3. As in [4], at each time step, the 1D profiles are first "spatially" processed by a sub-network consisting of $N_s$ convolutional layers. The output of this network contains a representation of the 1D features, as shown by the blue bar in Figure 1. The 1D features are then concatenated with the 0D scalar signals, and together form the complete input channels for the $N_t$ dilated convolutional layer blocks. Each of these convolutional layer blocks consists of dilated causal convolutional layers, activation layers, optional dropout layers, normalization layers, as well as an additive identity map (which maintains the stability of the neural network when it becomes "deep"). Details of the convolutional layer blocks were introduced in [17].

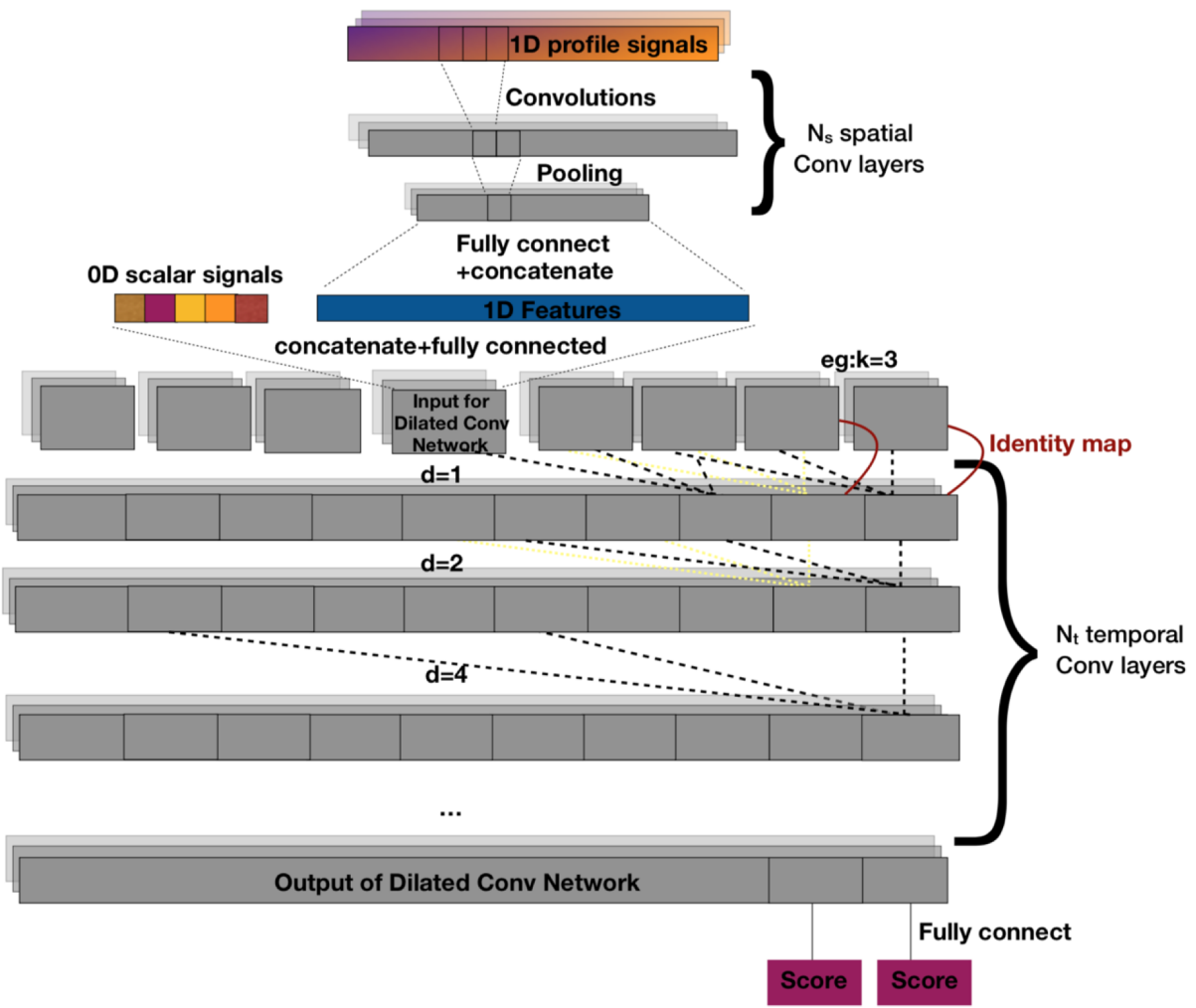

Fig. 1. The detailed schematic of our deep learning model based on temporal convolutional architecture. $N_s$ is the number of convolutional layers for spatial information processing at each time step, $N_t$ is the number of dilated temporal convolutional layers, d is the dilation factor, and k is the filter size for the dilated convolutional layers.

| Signal description | Numerical scale DIII-D | Numerical scale JET |
|---|---|---|
| Plasma current | 3.8 e-1 MA | 5.03 e-1 MA |
| Plasma current target | 3.9 e-1 MA | Not available on JET |
| Plasma current error | 3.1 e-2 MA | Not available on JET |
| Plasma current direction | 1.0 | Not available on JET |
| Internal Inductance | 2.02 e-1 | 1.51 e-1 |
| Plasma density | 1.19 e19 $m^{-3}$ | 4.69 e19 $m^{-3}$ |
| Input power (beam for DIII-D) | 1.85 e6 W | 4.47 e6 W |
| Radiated power core | 4.58 e2 W | 4.05 e4 W |
| Radiated power edge | 4.94 e2 W | 2.72 e4 W |
| Stored energy | 2.79 e5 J | 1.2 e6 J |
| Locked mode amplitude | 1.14 e-6 T | 5.72 e-5 T |
| Safety factor q95 | 1.0 | 1.0 |
| Normalized plasma pressure | 6.91 e-3 | Not available on JET |
| Input beam torque | 1.47 Nm | Not available on JET |
| Electron temperature profile | 9.53 e-1 keV | 1.53 keV |
| Electron density profile | 1.47 e19 m-3 | 2.98 e19 $m^{-3}$ |

Table 1. Measured DIII-D and JET experimental signals used in FRNN

Compared with the LSTM based models, TCN based models have two main advantages for processing the temporal information. First, instead of carrying over historical information in a recurrent fashion, the TCN directly fetches it through a time series, and can accordingly "remember" such information from a more distant past. How to effectively learn long-term dependencies for predictions involving time series databases is an active area of modern computational science research. While the LSTM is an improved architecture over the standard recurrent neural network (RNN) – which has short memory due to the exploding and vanishing gradient problems [18] – it can still lose distant information through operations on the cell state, which carries its long term memory. On the other hand, while the temporal receptive field of regular convolutions grows linearly with network depth (necessitating prohibitively deep architectures for long-term memory tasks), the temporal receptive field of dilated convolutions grows exponentially with network depth. TCNs can thus capture long distance dependencies with a modest number of layers.

The second advantage of the TCN architecture is that it is easily amenable to accelerated training via model parallelism. The serial implementation of the model is straightforward; i.e., in addition to fully connected layers and activation functions, only the convolution operation (applied to both the spatial and temporal representations) is required. The resulting feedforward network does not incorporate gated functions or recurrent connections. In contrast, the recurrent nature of the LSTM and other models based on RNN are generally hard to parallelize within training examples on the model level [17]. We highlight the fact that for our exemplar FES application of interest, the training and inference on long experimental runs with dense temporal measurements leads to a large memory footprint – thereby making the TCN's efficient model parallelism a particularly attractive feature in the context of high performance computing (HPC). Specifically, the TCN architecture can effectively utilize the hardware capabilities of modern leadership class supercomputing systems – a fact that will be demonstrated later in this paper.

With a chosen timestep of 1ms, and a typical shot duration on DIII-D of several seconds (tens of seconds on JET), the typical length of a time series in our data set is thousands to tens of thousands of steps. Since the effective history length of a normal convolutional layer is k-1, where k is the convolution filter size, regular convolutional neural networks have a receptive field linear in depth. By contrast, that of dilated convolutional neural networks, where each layer has effective history length of (k-1)*d, can be exponential if the dilation factor d is grown accordingly. The need for dilations (see dashed connections in Figure 1) arises if we need to allow the network output near the end of the shot to depend on early plasma behavior – and to be able to do so without requiring unwieldy network depth.

The output of the dilated convolutional layer blocks feeds into a final fully-connected layer that combines the information from all of the hidden units. It then outputs the disruption score which measures the likelihood of an imminent disruption at each time step. The definition of this disruption score, or the "target" that FRNN is trained on, is effectively a hyperparameter that can be tuned. A detailed explanation of the tradeoffs involved in the selection of the target function is provided in the "Methods" section of [4].

For a disruptive shot, if the disruption score rises above the pre-set "alarm threshold" before the "warning time" (the latest acceptable alarm time before the actual disruption), it would count as a true positive (TP) prediction. For a non-disruptive shot, if the disruption score rises above the "alarm threshold" at any time, it would count as a false positive (FP) prediction. By shifting the 'alarm threshold' from minus infinity (model predicts disruption for every shot) to infinity (model predicts no disruption for any shot), a receiver operating characteristic curve (ROC curve) is produced. We use the area under the curve (AUC) of the ROC curve on the test data prediction results as the metric to evaluate model performance. Although for actual tokamak operations, a fixed alarm threshold is required, for offline studies such as those introduced in this work, the AUC of the ROC can effectively represent the general model performance.

III. Training and Prediction Results

As discussed in Ref. [4], the data for this work comes from the DIII-D tokamak located at General Atomics, San Diego, CA [19], and the EUROFusion JET tokamak located at the Culham

Science Centre for Fusion Energy in the UK [20]. The data used in this work are subsets of the previously analyzed and published database in [4]. The DIII-D data is sampled from shot numbers ranging from 125500 to 168555, while the JET data are sampled from the carbon wall campaigns C23-C27b and the ITER-like wall campaigns C28-30. The different wall conditions in the two types of JET campaigns – as reported previously [4] – result in distinct plasma dynamics and disruption mechanisms. In this work, time step size for the signal data is chosen to be 1ms. Using the same data selection and preprocessing procedure as described in the Methods section of [4], we consider a shot to be valid within the database if and only if all of the relevant signals contain data for a period of time longer than the warning time. In order to assess model performance on DIII-D with a 1 second warning time, the DIII-D shots are required to have non-NAN and non-flat data in all signals for at least 1 second in order to be considered "valid" and to thus be added to the FRNN training/validation/testing dataset. This procedure resulted in slightly smaller numbers of shots in training, validation, and testing sets as listed in parenthesis in Table 2 when compared to the values listed in the Extended Data Table 2 of [4].

|  | Single machine | | | | Cross Machine | |
|---|---|---|---|---|---|---|
| Training (#shots) | DIII-D (1702) | | | JET-CW(1956) | DIII-D (2268) | JET-CW(1956) |
| Validation (#shots) | DIII-D (837) | | | JET-CW(962) | DIII-D (1117) | JET-CW(962) |
| Testing (#shots) | DIII-D (846) | | | JET-ILW(1133) | JET-ILW(1133) | DIII-D(846) |
| Warning time | 30ms | 0.2s | 1s | 30 ms | 30ms | 30ms |
| FRNN 0D-LSTM | 0.93 | 0.90 | 0.72 | **0.95** | 0.85 | **0.76** |
| FRNN 0D-TCN | 0.93 | 0.90 | 0.74 | 0.95 | **0.91** | 0.73 |
| FRNN 1D-LSTM | 0.93 | 0.89 | **0.80** |  | 0.84 |  |
| FRNN 1D-TCN | **0.93** | **0.91** | 0.79 |  | 0.86 |  |

Table 2. Prediction results. Performance of the best models (highlighted in bold) on the test datasets, measured as AUCs at the warning time before a disruption. Four FRNN models – trained with (1D) or without (0D) the 1D profiles, based on the LSTM or TCN architectures –

are compared when trained on DIII-D and JET shots. For single machine prediction tasks using the DIII-D database, we carried out hyperparameter tuning for three different warning times, 30ms, 0.2s and 1s, and reported the performance from the best model.

In Table 2 we report the prediction results from four distinct FRNN architectures across several different experimental databases. Specifically, these four schemes are either LSTM based or TCN based models, each trained with and without the 1D profile information. We tuned hyperparameters for each architecture to select the best performing model for each of the following 6 different tasks. These six tasks are as follows: (i-iii) prediction of DIII-D disruptions with 3 different warning time cutoffs using models trained on separate DIII-D data; (iv) prediction for DIII-D disruptions with 30ms warning time using models trained on JET carbon wall data; and (v-vi) predictions for JET ITER-like wall shots disruptions with 30ms warning time using models trained on JET carbon wall data, or on DIII-D data. Since the JET carbon wall datasets did not include profile information, there are no FRNN 1D results for models trained or tested with this data. During the hyperparameter tuning process for each of the twenty entries in Table 2, we randomly selected 40 sets of hyperparameters chosen from a reasonable range of possible values and then trained 40 models in parallel using the training and validation data. Main hyperparameters with their representative values for the TCN based model are summarized in Table 3. In total, 800 (20 x 40) models were trained. After training, we selected the best model based on its performance on the validation dataset. Finally, we examined the accuracy of each optimal model using the appropriate test dataset and summarized the test performance in Table 2.

| Hyperparameter | Explanation | Representative value |
|---|---|---|
| $\eta$ | Learning rate | 9.08 e-5 |
| $\gamma$ | Learning rate decay per epoch | 0.99 |
| $N_{batch}$ | Training batch size | |
| $T_{warning}$ | Warning time for target function, which becomes positive at $T_{warning}$ | 20 |
| Target | Type of target function | ttd (function linear in time to disruption) |
| $N_t$ | Number of causal temporal convolutional layers | 8 |
| $N_s$ | Number of spatial convolutional layers | 2 |
| $\lambda$ | Weighting factor for positive examples | 16 |
| $K_t$ | Size of temporal convolutional filters | 11 |
| $K_s$ | Size of spatial convolutional filters | 7 |
| $N_{Tstack}$ | Number of stacks of temporal convolutional blocks | 2 |
| $n_{tf}$ | Number of temporal convolutional filters | 60 |
| $n_{sf}$ | Number of spatial convolutional filters | 20 |
| Dropout | Dropout probability | 0.05 |

Table 3. Hyperparameters to be tuned for the TCN based FRNN model, explanations of the hyperparameter symbol, and representative well performing values.

Comparing results from the four schemes studied, the best performing model for each task is highlighted in bold in Table 2. From this round of hyperparameter tuning run, for most of the tasks, the TCN-based model in FRNN performs equally well or better than the original LSTM-based model. As an example, the ROC curves for the best performing models (based on TCN or LSTM) on the DIII-D test dataset with 0.2 second warning time and including 1D profiles are shown in Figure 2. The two models perform equally well for the low false positive rate regime, and the TCN-based model performs slightly better in the high true positive rate regime. As the high true positive rate regime is of greater importance to disruption prediction models for future machines that could not afford false negative results, this result demonstrates an example where considering multiple deep learning models can contribute to stronger predictive power for disruptions. Figure 3 shows an example DIII-D shot in the test dataset, and the output of these two models. While both models show some change in their disruption score shortly before the 1200ms mark, only the TCN model shows a sufficient change to trigger a correct alarm. Near the end of the shot, the TCN model also outputs a higher disruption score, although this alarm is within the 30ms warning time, and is considered "too late".

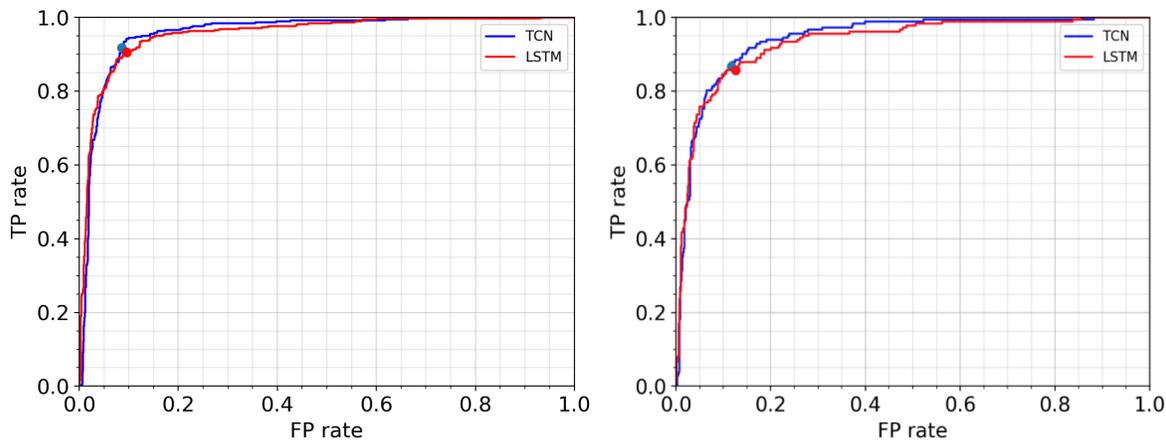

Fig 2. Comparison of ROC curves on the DIII-D training (left panel) and test (right panel) dataset with 0.2s warning time for the optimal FRNN 1D models, based on the TCN (blue) and LSTM (red) architectures. The solid dots indicate model performance at the optimal alarm threshold determined on the validation set. Both models demonstrate good generalization of the optimal alarm threshold from DIII-D validation data to DIII-D test data.

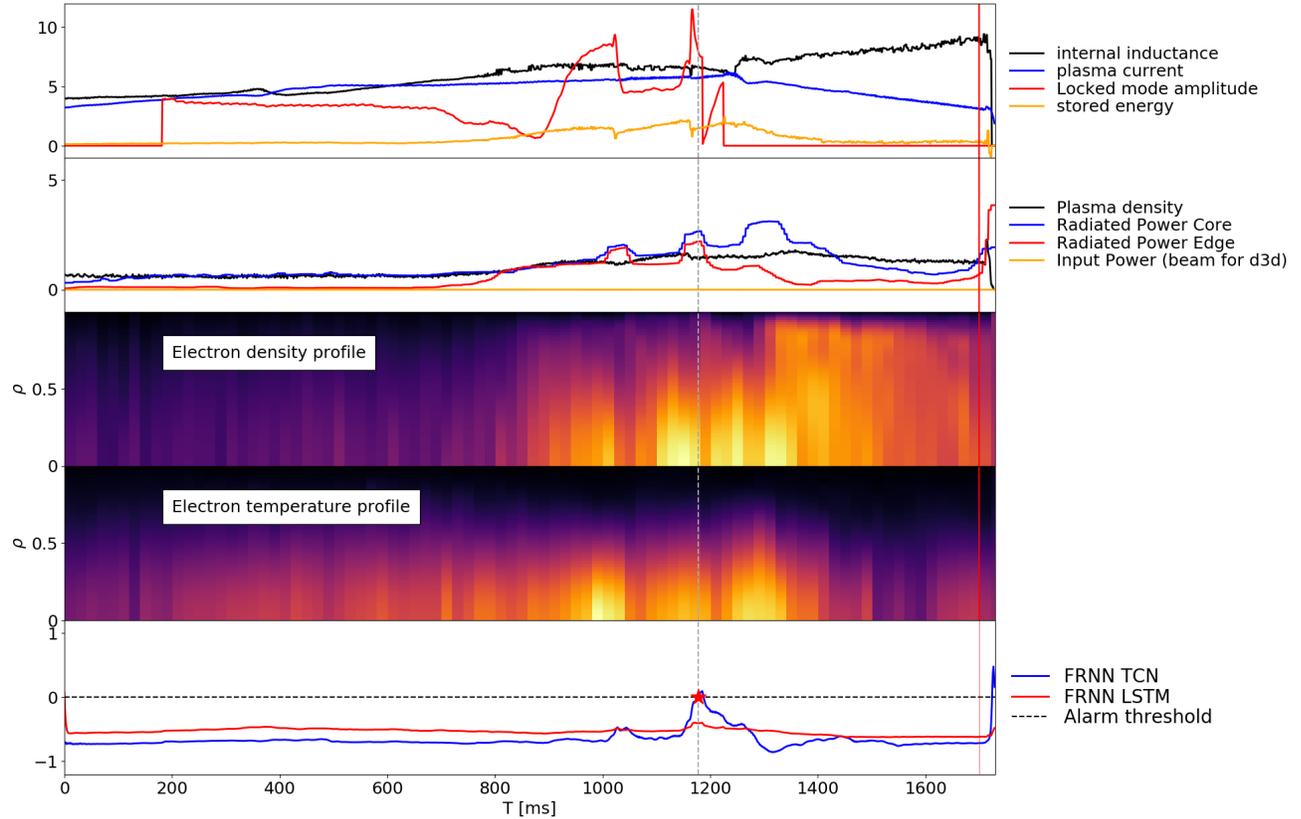

Fig 3. Example prediction on DIII-D shot # 147206. The top two panels show eight representative 0D scalar signal channels (normalized by numerical scale on DIII-D as listed in Table 1), and the next two panels show two 1D profile signals channels as FRNN inputs. Vertical axis of the 1D profile signals are the normalized toroidal magnetic flux $\rho$. The last panel shows the disruption scores returned by FRNN using TCN (blue line) and LSTM (red line) based models. The FRNN TCN model raised positive disruption alarm before 1200 ms when reaching disruption alarm threshold indicated by the horizontal dashed line. The signals are plotted up to the time of disruption. The solid vertical red line shows the latest warning time (30ms before the disruption). It is important to highlight the finding here that while both models respond noticeably around the indicated disruption alarm time, only the TCN based model correctly triggers the disruption alarm around 0.5 second before the actual disruption.

For the cross-machine prediction task where the model was trained only on DIII-D shots and is tasked to predict disruptions in the JET ITER-like wall shots, the TCN architecture offers significantly improved accuracy with less overfitting. The problem of overfitting is hard to overcome for this cross-machine predictive task due to drastically different physical parameters for DIII-D and JET plasmas. A comparison of the best TCN based model and the best LSTM based model performance on the cross-machine prediction task is shown in Figure 4. Although the LSTM based model demonstrates better performance on the training data from DIII-D plasmas, as shown in the left panel of Figure 4, the TCN based model achieves significantly better inference result for the JET data, as shown in the right panel of Figure 4. The optimal alarm threshold estimated base on the DIII-D validation data also generalizes much better to the JET test data for the TCN based model. This result indicates that the TCN based model has learned deeper physics-based information, which is general for both plasma devices, and can thus achieve disruption predictions for plasma conditions that it has never seen during training.

Figure 5 shows the signals from an example JET shot in the top four panels, and outputs from the FRNN models trained on DIII-D database in the last panel, where only the TCN based model is able to catch the continuously rising core radiation power as plotted in black line in the second panel, and predict for the imminent disruption, even when the input power is decreased, as plotted in green line in the second panel. On the contrary, the LSTM based model output had a sharp rise in disruption score at around 12s, when the input power is suddenly increased.

Although in the two aforementioned tasks, the TCN based models outperform the LSTM based models in FRNN, we highlight here that the LSTM based modes can outperform the TCN based models for other tasks, such as the prediction of disruptions in DIII-D plasmas when trained based on JET plasma signals, as shown in the last column in Table 2. It is therefore important to recognize that different types of deep learning architectures can be suitable for different types of physics problems, and even different facets for the same problem, and the possible generally improved predictive powers if the ensemble model based on multiple diverse deep learning architectures are considered, especially for cross-machine predictive tasks.

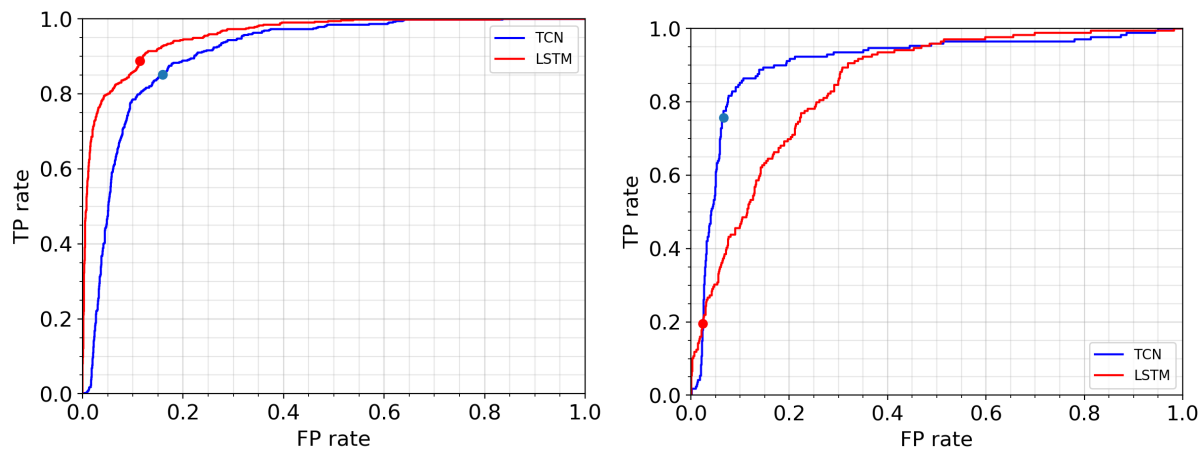

Fig 4. Comparison of ROC curves on the DIII-D training (left panel) and JET test (right panel) dataset for the optimal FRNN 0D models, based on the TCN (blue) and LSTM (red) architectures. The training and test score for the LSTM based model is 0.96 and 0.85 respectively. For the TCN based model, both training and test score is 0.91, indicating much less overfitting. The solid dots indicate model performance at the optimal alarm threshold determined on the validation set. The generalization of the optimal alarm threshold from DIII-D validation set to JET test set is significantly better.

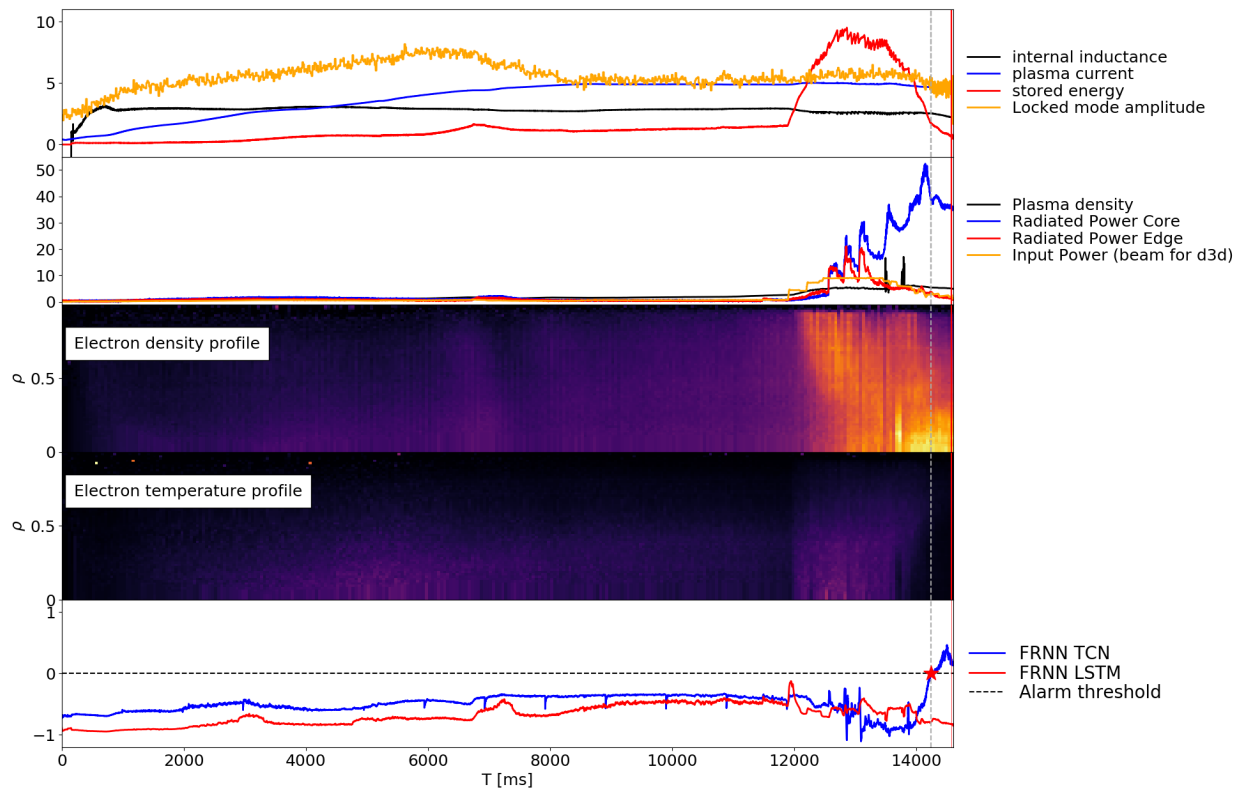

Fig 5. Example prediction on JET shot #83340. The top two panels show eight representative 0D scalar signal channels (normalized by numerical scale on JET as listed in Table 1), and the next two panels show two 1D profile signals channels as FRNN inputs. The last panel shows the disruption scores returned by FRNN using TCN (blue line) and LSTM (red line) based models. The signals are plotted up to the time of disruption. The dashed vertical line indicates the disruption alarm time, and the solid vertical red line shows the latest warning time (30ms before the disruption).

In order to achieve an improved level of understanding of the physics and signal features underlying these disruption prediction results, we carried out a series of sensitivity studies. Specifically, we assessed the signal importance of all input signals for the single machine disruption prediction task on DIII-D using FRNN-1D – similar to what was carried out in earlier work [4]. To assess the contribution from each of the 16 signals to the disruption scores, we trained 16 individual models each using one single signal at a time. In Figure 6, test performances are compared for the FRNN-1D TCN-based (left panel) and the LSTM-based (right panel) models trained with the single labeled signal. As expected, for both architectural schemes, the models trained with single channels have significantly lower performance than the model trained with all 16 signals (represented by the deep blue bar). The signal importance as measured by model performance for each of the two models are affected by multiple factors, including initialization stochasticity and model hyperparameters. Therefore, some variation in signal importance values between the models is expected. However, it is important to note that there are clear qualitative trends that are consistent in both panels of Figure 6. For example, the models trained on either the locked mode amplitude (LM) or the tokamak safety factor value approaching the plasma periphery (q95) signals outperform models trained on the rest of the signals, indicating that these two signals contain key disruption related information. Moreover, the trend in this sensitivity study of signal importance also aligns well with that in the extended data Fig. 2 (a) in [4], where signal importance results for an LSTM-based model were reported using DIII-D data.

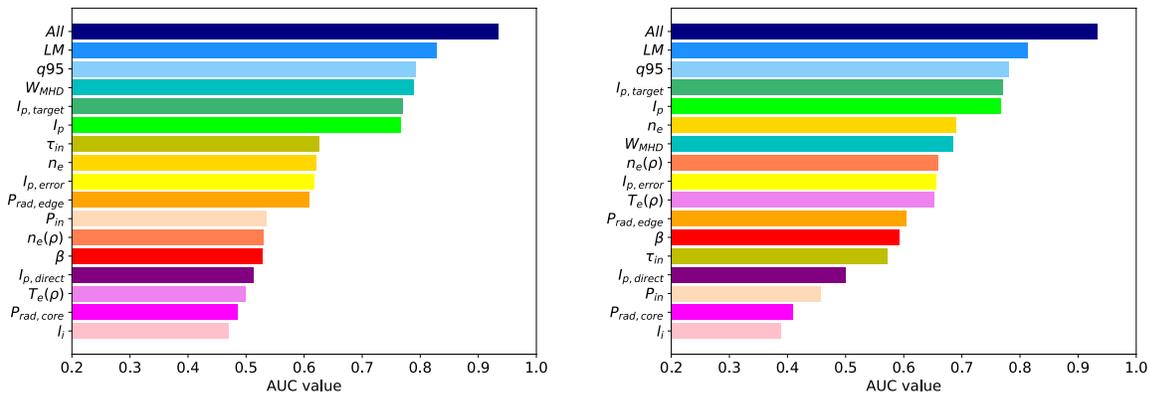

Fig 6. Signal importance studies for models based on the TCN (left) and LSTM (right) architectures. Each bar represents the test set AUC values achieved by a model trained on the single labeled signal. The general trend of this sensitivity study is similar for the two architectures, showcasing the robustness of this method in estimating the relative importance of different physical signals.

In each panel in Figure 6, all models are trained using the same set of hyperparameters as those used for the respective best model tuned with all signals (represented by the deep blue bar), and 34 models in total are trained for this analysis. In future investigations, more reliable estimates could be obtained by running hyperparameter tuning for each of these models, making it necessary to train thousands of models. Such a task would clearly require the associated engagement and effective utilization of very powerful modern supercomputers.

IV. Computational Performance Evaluation

As demonstrated in this paper, the training and tuning of deep learning computational software like FRNN requires modern supercomputing power that must be utilized with excellent efficiency, as was also discussed in earlier work [4]. Here, we demonstrate that with the new TCN architecture, FRNN exhibits even better computational performance. Specifically, Figure 7 shows performance comparisons for the best TCN-based and the best LSTM-based models, when both are trained on the DIII-D database with 30 ms warning time. For this task, the key finding is that the TCN architecture enabled FRNN to reduce training time by about a factor of 2.

As shown in Figure 8, FRNN exhibits impressive strong scaling on the Oak Ridge Leadership Computing Facility (OLCF) "Summit" – currently the top-rated supercomputer worldwide. Specifically, the training time for this TCN-based model scales almost ideally with the number of GPUs used. In this scaling study, we used a much larger database and a different set of model hyperparameters than that used in Figure 7, to avoid MPI operation problems when the training time becomes too small for a large number of GPUs. This important finding motivates the future development of modern deep learning software such as FRNN in various aspects – including performance optimization via hyperparameter tuning, consideration of more complicated and deeper architectures, and the addition of higher dimensional input data sources with higher resolution.

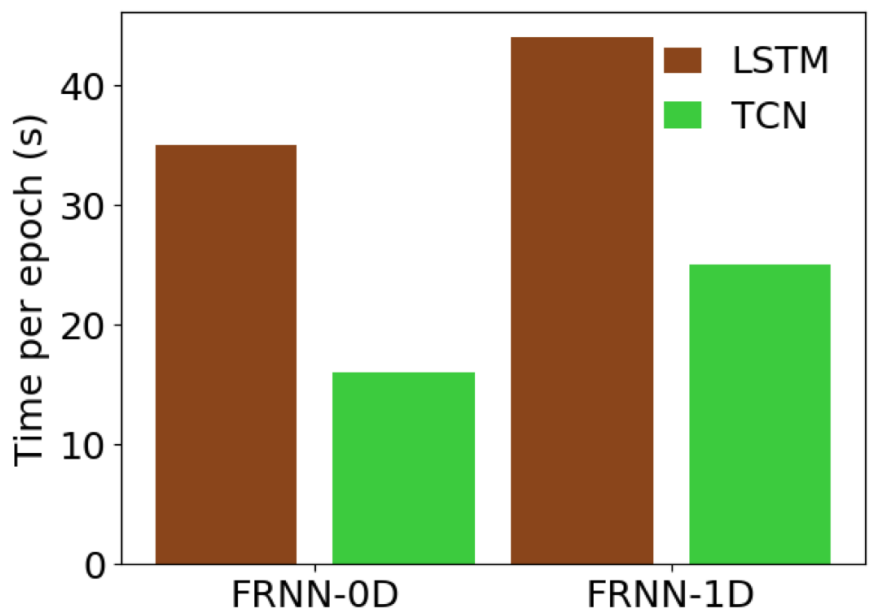

Fig 7. Time per epoch (i.e., the time required to complete one pass over the entire training dataset) during training using 4 Tesla V100s for FRNN-0D and FRNN-1D, for the LSTM (brown) and TCN (green) architectures, respectively. Lower values correspond to better computational performance. The four models studied here correspond to the best models from the studies of DIII-D single machine disruption prediction with 30ms warning time (see first column in Table 2).

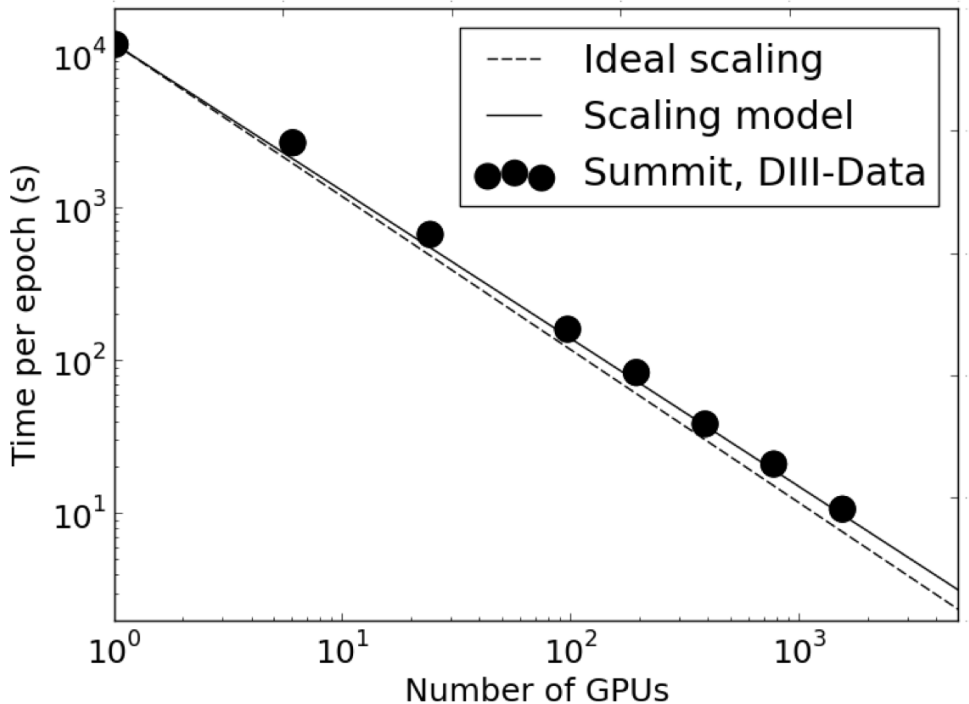

Fig 8. Strong scaling of FRNN with the new TCN-based model carried out on the Oak Ridge Leadership Computing Facility (OLCF) Summit system – currently the #1 rated supercomputer worldwide [21]. The time required to complete one epoch during training on Summit (black data points) agrees well with the scaling model. The original strong scaling of FRNN using the LSTM-based model and the associated derivation of the scaling model can be found in [4].

V. Summary and future work

In this paper, we have introduced and implemented a new architectural scheme based on dilated temporal convolutions within the exemplar magnetic fusion plasma disruption prediction software FRNN [4]. Compared with the previously published models based on the LSTM architecture, FRNN models constructed with the temporal convolutional neural network (TCN) architecture exhibit at least equivalent and demonstrably superior computational performance and predictive power for disruption forecasting across various experimental databases from the DIII-D and JET tokamaks. The TCN architecture has also been applied in other disruption studies recently, using input from the Electron Cyclotron Emission imaging (ECEi) diagnostic data on DIII-D [22].

In the present paper, we have developed a general deep learning capability to train multiple models with distinct architectures within a single software suite that serves to promote adaptability to different temporal and spatial learning tasks and also to enable ensemble schemes for highly accurate prediction. Various machine learning based algorithms targeting different learning and prediction tasks have been independently validated for modern tokamaks such as DIII-D [4, 14] and JET [23-35]. The capability of utilizing effective ensemble models in real-time plasma control systems is an important task for successful operation of future machines. To contribute to this effort, we plan to implement additional deep learning based architectures, including those based on attention-based models (such as the "Transformer" [36]), into the FRNN framework, as well as utilizing more high dimensional experimental data from diagnostics including ECEi and Thomson Scattering. The goal is to develop FRNN into a flexible and adaptable software platform for the prediction and analysis of complex plasma dynamics, including early disruption prediction as well as other important physics phenomena. Beyond plasma device performance predictions, this platform also has potential for wide applications to time series prediction problems in other research areas, such as magnetosphere substorm onset predictions and weather forecasts.

Acknowledgements

Research for this paper was carried out at the Princeton Plasma Physics Laboratory by the Department of Energy (DOE) contract DE-AC02-09CH11466. The authors thank Professor Zhihong Lin of UC Irvine for useful discussions and important support of this work associated with the (DOE) SciDAC ISEP Center which he leads. We have benefited from the HPC resources of the Oak Ridge Leadership Computing Facility at the Oak Ridge National Laboratory (DOE Contract No. DE-AC05-00OR22725) and the National Energy Research Scientific Computing Center (DOE Contract No. DE-AC02-05CH11231. This work is also supported under Contract DE-AC02-06CH11357 associated with the Argonne Leadership Computing Facility (ALCF) Aurora Early Science Program project at the Argonne National Laboratory. The simulations presented in this article were performed partly on computational resources featuring the "Traverse" cluster managed and supported by Princeton University's Research Computing Center, a consortium of groups including the Princeton Institute for Computational Science and Engineering (PICSciE) and the Office of Information Technology (OIT). We also express our gratitude to the EUROfusion Joint European Torus (JET) and their management as well as to General Atomics (GA) and its DIII-D tokamak project for access to the same fusion databases which were previously provided for the Nature (April, 2019) publication. In addition, we extend special thanks to Dr. Nik Logan of PPPL for his careful reading and associated helpful suggestions that improved the clarity of this manuscript. This material is based upon work supported by the US DOE, Office of Science,
Office of Fusion Energy Sciences, using the DIII-D National Fusion Facility, a DOE Office of Science user facility, under award DE-FC02-04ER54698.

product, or process disclosed, or represents that its use would not infringe privately owned rights. Reference herein to any specific commercial product, process, or service by trade name, trademark, manufacturer, or otherwise does not necessarily constitute or imply its endorsement, recommendation, or favoring by the United States Government or any agency thereof. The views and opinions of authors expressed herein do not necessarily state or reflect those of the United States Government or any agency thereof.